\documentclass[aps,pre,showpacs,amsmath,amsfonts,amssymb,superscriptaddress,preprint]{revtex4-1}

\usepackage{graphicx,multirow}
\usepackage{amsthm}
\usepackage{framed}
\usepackage{amssymb}
\usepackage{amsmath,mathrsfs}
\usepackage{color}
\usepackage{array}
\usepackage{lipsum}
\usepackage[table]{xcolor}

\usepackage{sepfootnotes, longtable}
\newcommand{\affA}{Aix-Marseille University, Marseille, France}
\newcommand{\affB}{CNRS Centre de Physique Th\'eorique UMR7332,
13288 Marseille, France}
\newcommand{\affC}{ Department of Chemistry, Rice University, Houston, USA}
\newcommand{\affD}{Department of Oncology, 3-336, Cross Cancer Institute, 
Edmonton, AB, T6G 1Z2, Canada  }
\newcommand{\affE}{ Department of Physics, University of Alberta, Edmonton, Canada. }

\definecolor{Gray}{gray}{0.9}
\newcolumntype{g}{>{\columncolor{Gray}}c}
\everymath{\displaystyle}

\begin{document}

\DeclareGraphicsExtensions{.jpg,.pdf,.mps,.png}

\title{On the role of electrodynamic interactions in long-distance biomolecular recognition}

\author{Jordane Preto}
\email{preto@cpt.univ-mrs.fr} 
\affiliation{\affC}
\affiliation{\affA}
\affiliation{\affB}
\author{Marco Pettini}
\email{pettini@cpt.univ-mrs.fr}
\affiliation{\affA}\affiliation{\affB}
\author{Jack A. Tuszynski}
\email{jack_BT@ualberta.ca}
\affiliation{\affD}\affiliation{\affE}

\begin{abstract}

The issue of retarded long-range resonant interactions between two molecules with oscillating dipole moments is reinvestigated within the framework of classical electrodynamics. By taking advantage of a theorem in complex analysis, we present a simple method to calculate the frequencies of the normal modes, which are then used to estimate the interaction potential. The main results thus found are in perfect agreement with several results obtained from quantum computations. Moreover, when applied in a biophysical context, our findings shed new light on Fr\"ohlich's theory of selective long-range interactions between biomolecules. In particular, at variance with a long-standing belief, we show that sizeable resonant long-range interactions may exist only if the interacting system is out of thermal equilibrium.

\end{abstract}

\date{\today}


\maketitle

\section{Introduction}

For several decades now, understanding biomolecular dynamics and general mechanisms of protein-protein or protein-DNA recognition has become a diverse and fascinating topic which still generates much interest in the biophysics community. Not only does this area of research seem unavoidable to explain the characteristic synergy within most of biological organisms but it can also be applied \textit{e.g.}, to investigate the microscopic origin of infectious diseases as well as to the design of new drugs whose efficiency depends on how fast the synthesized proteins find their biological target \cite{zhao}. When dealing with biomolecular recognition, it is suitable to distinguish between the phase during which (globular) proteins need to find their target while freely moving within the aqueous biological environment (3D diffusion) and the phase when the protein and its target, which have finally arrived in each other's vicinity, are in close contact. The latter is characterized by orientational and conformational adjustments of the biomolecules so that a biochemical reaction can eventually take place between them. At this stage, molecular motions are mainly governed by chemical forces of short-range nature (hydrophilic/hydrophobic interactions, van der Waals forces, covalent bonds, etc.) besides the traditional Brownian motion due to thermal fluctuations. Accurate information regarding the electrostatic charge distribution is needed to describe the dynamics and to estimate the associated rate constants.  As far as the run-up to molecular encounter is concerned, the detailed topology of the molecules involved becomes irrelevant as the protein and its target are supposed to be far enough from one another. From a point of view of the dynamics, intermolecular electrostatic interactions might \textit{a priori} play a role as those interactions are inherently of a long-range kind. However, cell cytoplasm contains considerable amounts of small ions which tend to screen any static electric charges at short range. The Debye length is typically smaller than 10$\mathrm{\AA}$ in cellular media. Even though the electrostatic field is screened at short distance, Debye screening turns out to be ineffective for electric fields oscillating at large enough frequencies. Xammar Oro showed experimentally that an electrolyte such as cell cytoplasm behaves like a pure dielectric (\textit{i.e.}, free of conducting properties) when acted upon by an electric field with frequencies larger than 250 MHz \cite{xammar}. Such a threshold frequency was originally identified by Maxwell through analytical arguments in his "Treatise on Electricity and Magnetism" \cite{maxwell}. In any case, this suggests that electrodynamic intermolecular forces, \textit{i.e.}, forces between molecules carrying oscillating charges, might influence the encounter dynamics of biological cognate partners. Further theoretical investigations of these forces are required to probe their real contribution in a cellular environment.

Long-distance electrodynamic forces between two neutral atoms or small molecules have been broadly investigated in quantum electrodynamics (QED) \cite{qed}. Those forces typically arise when one of the atoms is in an excited state, and the transition frequencies of both atoms are similar. In quantum terms, this corresponds to the so-called exchange degeneracy which requires that the atoms remain in a quantum entangled state so that a net attraction (or repulsion) takes place \cite{qed}. Entangled states are very fragile, and, especially in a biological context, their existence over long distances could be questioned because of the noisy cellular environment. Thus, it is worth investigating long-distance interactions between biomolecules on a classical electrodynamics basis. 

In the present paper, we will show that, similarly to QED, long-distance electrodynamic interactions can be activated in conditions of classical resonance (off-resonance conditions would lead to short-distance interactions). Numerical estimates of resonant and non-resonant forces will be given for typical biomolecules. While quantum electrodynamic interactions are attributed to instantaneous dipole moments resulting from electronic transitions of the atoms, dipole moments involved in the classical limit are due to conformational molecular vibrations. Since already three decades ago, experimental evidence is available regarding the existence of low-frequency excitations of macromolecules of biological relevance (proteins \cite{proteins} and polynucleotides \cite{nucleotides}) through the observation of the Raman and Far-Infrared spectra of polar molecules. These spectral features are commonly attributed to coherent collective oscillation modes of the whole molecule (protein or DNA) or of a substantial fraction of its atoms. These collective conformational vibrations bring about oscillations of the total electric dipole moment, suggesting that beyond all the well-known short-range forces, biomolecules could interact also at a long distance by means of electrodynamic forces, provided they undergo these collective excitations. We think of collective excitations as they \textit{a priori} give rise to large dipole moments, they can be switched on and off by suitable environmental conditions, and, finally, because they can induce strong resonant dipole interactions between biomolecules when they oscillate at the same frequency or with the same pattern of frequencies. Resonance would thus result in selective forces. These collective macromolecular vibrations are observed in the frequency range of 0.1-10 THz.
In other words, there are reasons based on physical first principles to assign an essential role to an electromagnetic communication among biomolecules to account for the aforementioned highly efficient pattern of biochemical reactions in cells. This possibility has been hitherto overlooked, mainly because of two reasons. On the one hand, in the framework of short-range shape or dynamic complementarity (lock and key and induced-fit) between cognate molecular partners, diffusion driven random encounters between biomolecules seemed a natural and sufficient explanation. On the other hand, even though the idea of long-distance electromagnetic interactions between macromolecules has been sometimes surmised by physicists, the absence of any experimental strategy to convincingly detect their actual activation has marginalized this hypothesis. In this respect this work is motivated by the present day feasibility of experimental tests to assess whether such interactions could be relevant at the biomolecular level \cite{firstpaper,jthesis,ithesis}. Hence the need for a thorough revisitation of the theoretical framework.

Based on apparently standard computations of classical electrodynamics,  our present work resorts to a powerful inversion theorem in complex 
analysis and results in non-trivial development of a fascinating theory pioneered by H. Fr\"ohlich \cite{frohlich72,frohlich78,frohlich80} shedding new light on some of its former results. In particular, we show that a commonly accepted result at thermal equilibrium is incorrect, that is, an electrodynamic interaction potential proportional to $-1/r^3$ (with $r$ the intermolecular distance) can be activated only out of thermal equilibrium. Moreover,  additional interaction terms proportional to $-1/r^2$ and $-1/r$, both modulated in space, are found as field retardation effects; such terms are well known in QED, though the spatial modulation is still controversial, but in our classical framework these interaction terms are not associated with entanglement, as is the case with QED. A preliminary account of some of the results reported here was given in Ref. \cite{pla}.

\section{Electrodynamic Interactions in Classical Physics}\label{classical interactions}

In order to assess  whether electrodynamic interactions may play a sizeable role in the organization of biomolecular reactions, 
in particular, by facilitating encounters of biomolecular cognate partners over long distances, the following sections are devoted to the
investigation of classical electrodynamic interactions  between two oscillating molecular dipoles. We focus on resonant properties 
of these interactions so that a particular biomolecule would be only attracted by its specific target, and not by other neighboring biomolecules.
Field retardation effects are also taken into account.

\subsection{Equations of motion}

The far-field electrodynamic interactions between molecular systems mainly involve their dipole moments, that is, multipolar
contributions can be neglected. Hence,  we consider a simple system of two molecules $A$ and $B$ with dipole moments
$\boldsymbol{\mu}_A$ and $\boldsymbol{\mu}_B$ oscillating with harmonic frequencies $\omega_A$ and $\omega_B$, respectively.
The corresponding equations of motion are written in the following general form 

\begin{equation}\label{system0}
\left \{
\begin{array}{l}
 \ddot{\boldsymbol{\mu}}_A + \gamma_A \dot{\boldsymbol{\mu}}_A + \omega_A^2 \boldsymbol{\mu}_A =
\zeta_A\boldsymbol{E}_B(\boldsymbol{r}_A,t) + \boldsymbol{f}_A(\boldsymbol{\mu}_A,t) \vspace{3mm} \\
 \ddot{\boldsymbol{\mu}}_B  + \gamma_B \dot{\boldsymbol{\mu}}_B +  \omega_B^2 \boldsymbol{\mu}_B =
\zeta_B \boldsymbol{E}_A(\boldsymbol{r}_B,t) + \boldsymbol{f}_B(\boldsymbol{\mu}_B,t).
\end{array}
\right.
\end{equation}

\medskip

Since we have adopted the dipole approximation, the interaction
between molecules is mediated by the electric field $\boldsymbol{E}_{A, B}(\boldsymbol{r},t)$ created by each dipole,
here located at $\boldsymbol{r}=\boldsymbol{r}_{B,A}$.
Related coupling constants are $\zeta_{A} = Q_{A}^2/m_{A}$,
with $Q_A$ and $m_A$ the effective charge and mass of dipole $A$, and $\zeta_B$ with a similar $B$-labeled expression. Other quantities
are the damping coefficients $\gamma_{A,B}$  of the dipoles representing radiation losses, and functions $\boldsymbol{f}_{A,B}$
that describe, from a general point of view, possible anharmonic contributions of each dipole as well as possible external excitations.

Our goal here is to estimate the mean interaction energy of the system given by Eqs. \eqref{system0}. To proceed,
we calculate its normal frequencies; starting from the associated harmonic conservative system:

\begin{equation}\label{system}
\left \{
\begin{array}{l}
 \ddot{\boldsymbol{\mu}}_A + \omega_A^2 \boldsymbol{\mu}_A = \zeta_A\boldsymbol{E}_B(\boldsymbol{r}_A,t) \vspace{3mm} \\
 \ddot{\boldsymbol{\mu}}_B  + \omega_B^2 \boldsymbol{\mu}_B = \zeta_B \boldsymbol{E}_A(\boldsymbol{r}_B,t),\\
\end{array}
\right.
\end{equation}

the normal frequencies are defined as the frequencies $\omega_N$ such that
$
\boldsymbol{\mu}_{A,B}(t) = \boldsymbol{\mu}_{A,B} e^{i \omega_N t}
$ are solutions of Eqs. \eqref{system}.  In order to get $\omega_N$, expressions of $\boldsymbol{E}_{A,B}(r,t)$ are computed explicitly.
The computation of the electromagnetic field generated by an oscillating dipole is detailed in the Appendix. To
link up results of the Appendix to our present problem, we assume that the dipole moment $\boldsymbol{\mu}$ in Eq. \eqref{electromag} oscillates harmonically at frequency
$\omega_N$, \textit{i.e.}, $\boldsymbol{\mu} = \boldsymbol{\mu}_{A,B} \delta(\omega - \omega_N)$. Substituting this relation into Eq. \eqref{electromag},
we get after inverse Fourier transform:
\begin{equation}\label{field}
 \boldsymbol{E}_B(\boldsymbol{r}_A, t) = \boldsymbol{\chi}(r,\omega_N)\boldsymbol{\mu}_{B} e^{i \omega_N t},
\end{equation}

with $r =|\boldsymbol{r}_A - \boldsymbol{r}_B |$ the intermolecular distance; an analogous expression is given for $\boldsymbol{E}_A(\boldsymbol{r}_B, t)$.
Here $\boldsymbol{\chi}(r,\omega)$ represents
the susceptibility matrix of the electric field and is derived in the Appendix [Eq. \eqref{electricsus0}]. For $\omega \in \mathbb{C}$, one has

\begin{subequations}\label{electricsus}
\begin{gather}
\begin{array}{l}
\displaystyle \chi_{11}(r,\omega) = \chi_{22}(r,\omega)  = - \frac{e^{\pm i \omega \sqrt{\varepsilon(\omega)} r /c}}{\varepsilon(\omega) r^3} \cdot
\vspace{0.3cm}\\
\hspace{2.5cm}
\displaystyle \left(  1 \mp  \frac{ i\omega
\sqrt{\varepsilon(\omega)} r}{c}  -    \frac{\omega^2
\varepsilon(\omega) r^2}{c^2}  \right), \vspace{0.3cm}
\end{array} \\
\chi_{33}(r,\omega)  =  \frac{2e^{\pm i \omega \sqrt{\varepsilon(\omega)} r /c}}{\varepsilon(\omega) r^3}
\left(  1 \mp \frac{ i\omega
\sqrt{\varepsilon(\omega)} r}{c}   \right),
\end{gather}
\end{subequations}

with $\chi_{ij}(r,\omega)  =  0$ when $i \neq j$; the $\pm$ sign is attributed to positive or negative values of
$\text{Im}(\omega\sqrt{\varepsilon(\omega)})$, respectively.
We remark again that the diagonal form of $\boldsymbol{\chi}$ is due to
the choice to set the $z$ axis along $r$. For real values of $\omega_N$, each element
$\chi_{ii}(r,\omega_N)$ of $\boldsymbol{\chi}(r,\omega_N)$ is a
complex number whose imaginary part accounts for the dissipation due to the field propagation \cite{landaubis}.  Then, since in
computing normal frequencies  one drops dissipation effects,
only the real parts of each element of $\boldsymbol{\chi}$, denoted by $\chi'_{ii}$, are considered in what follows.

\medskip

By substituting into Eq. \eqref{system} the assumed harmonic forms for $\boldsymbol{\mu}_{A,B}(t)$ and using
Eq. \eqref{field}, one obtains

\begin{equation}\label{systembis}
\left \{
\begin{array}{l}
 (\omega_A^2 - \omega_N^2) \mu_{A,i} = \zeta_A\chi_{ii}'(r,\omega_N) \mu_{B,i}  \vspace{0.3cm} \\
(\omega_B^2 - \omega_N^2) \mu_{B,i} = \zeta_B \chi_{ii}'(r,\omega_N)\mu_{A,i} ,  \\
\end{array}
\right.
\end{equation}

\medskip

The existence of solutions of the system \eqref{systembis} is ensured by the vanishing of its determinant, \textit{i.e.},
$(\omega_A^2 - \omega_N^2)(\omega_B^2 - \omega_N^2) -  \zeta_A\zeta_B (\chi_{ii}'(r,\omega_N))^2  = 0$. After trivial algebra, we get for each $i$
two possible solutions for $\omega_N^2$ that we call $\omega_{i,+}^2$ and $\omega_{i,-}^2$; these satisfy

\begin{equation}\label{secdegre}
\omega_{i,\pm}^{2} -  \frac{1}{2} \left\{ \left( \omega_A^2 + \omega_B^2\right) \pm \sqrt{\left( \omega_A^2 - \omega_B^2\right)^2
+ 4 \zeta_A\zeta_B (\chi'_{ii}(r,\omega_{i,\pm}))^2} \right\} = 0.
\end{equation}

\medskip

By computing the normal frequencies $\omega_{i,\pm}$ of the system \eqref{system}, we can rewrite it as a system
of six uncoupled harmonic oscillators of frequencies $\omega_{i,\pm}$, $i=1,2,3$ and energies $E_{i,\pm} = \omega_{i,\pm} J_{i,\pm}$, with
$J_{i,\pm}$ the action constants depending on initial conditions. In Eq. \eqref{secdegre}, $\omega_{i,\pm}$ is given in a
complicate implicit way. Explicit approximate expressions for $\omega_{i,\pm}$
are needed to compute the energy and then the interaction energy of the system of dipoles. In particular, we \text{will see}
that \textit{the range} of the interaction is strongly dependent on whether the dipoles oscillate at similar frequencies or not.

\subsection{Off-resonance Case}

When $\omega_A \gg \omega_B$ (or  when $\omega_A \ll \omega_B$), Eq. \eqref{secdegre}, for all $i$, becomes, at lowest order,

$$
\omega_{i,\pm}^{2} -  \frac{1}{2} \left\{ \left( \omega_A^2 + \omega_B^2\right) \pm \left( \omega_A^2 - \omega_B^2\right) \left( 1
+ \frac{2\zeta_A\zeta_B (\chi'_{ii}(r,\omega_{i, \pm}))^2}{\left( \omega_A^2 - \omega_B^2 \right)^2 } \right) \right\} \simeq 0,
$$

\medskip

which leads to :

\begin{equation} \label{nonresonance}
\begin{array}{c}
\underbrace{\omega_{i,\pm}^2 -  \omega_{A,B}^2  \mp  \frac{ \zeta_A\zeta_B(\chi'_{ii}(r,\omega_{i, \pm}))^2}{ \omega_A^2 - \omega_B^2 }}
_{\displaystyle \Theta_{i,\pm}(r,\omega_{i,\pm})}  = 0 \ , \ \
\text{where} \  \  \omega_{A,B} \ \  \text{means} \ \left\{ \begin{array}{l}
						  \omega_A \ \ \text{for} \ \  \omega_{i,+}, \vspace{0.3cm} \\
						   \omega_B \ \ \text{for} \ \ \omega_{i,-}.
						   \end{array} \right.

\end{array}
\end{equation}

Now, to find explicit solutions of Eq. \eqref{nonresonance}, we apply the Lagrange inversion theorem of complex analysis 
which states that  \cite{whittaker}:

\bigskip

{\it
 Let $\mathscr{C}$ be a contour in the complex plane, and let
$f$ a function analytic inside, and on $\mathscr{C}$. Let $\Theta$ another function which is analytic inside it 
and on $\mathscr{C}$ except at a finite number of poles. Noting $a_1,...,a_n$
the zeros of $f$ in the interior of $\mathscr{C}$, of degree of multiplicity $r_1,..r_n$, and $b_1,...,b_m$
the poles of $f$ of degree of multiplicity $s_1,...,s_m$, one has the following formula:

\begin{equation}
\sum \limits_{j=1}^n r_j f(a_j) - \sum \limits_{k=1}^m s_k f(b_k) = \frac{1}{2i \pi}
\oint\limits_\mathscr{C} dz f(z) \partial_z \ln \Theta(z).
\end{equation}
}

In the present context, if $\mathscr{C}$ is a contour on and inside which $\Theta_{i,\pm}$ is analytic, and that contains only
one solution $\omega_{i,\pm}(r)$ of Eq. \eqref{nonresonance}, $f$ can be taken as the identity function and $\Theta$ as
$\Theta_{i,\pm}(r,\omega)$, so that:

\begin{equation}\label{argument}
\omega_{i,\pm}(r) =  \frac{1}{2 i \pi } \oint \limits_{\displaystyle \mathscr{C}}  dz \ z \ \partial_z \ln \Theta_{i,\pm}(r,z) .
\end{equation}

Now, regarding the contour, we choose $\mathscr{C}$ on the right part of the complex plane (complex numbers
with positive real values) and we suppose that the inequality

\begin{equation}\label{ineq}
\frac{ \zeta_A\zeta_B(\chi'_{ii}(r,z))^2}{ \omega_A^2 - \omega_B^2 } <
\left| z^2 -  \omega_{A,B}^2 \right|,
\end{equation}

holds for all $z$ on the perimeter of $\mathscr{C}$. On the other hand, since each $\chi'_{ii}$, $i=1,2,3$, is a sum of inverse power
laws of $r$, we can assume that, for large $r$, relation \eqref{ineq} is satisfied. In this case, by applying Rouch\'e's theorem,
it is seen that $\omega_{A,B}$ is located inside $\mathscr{C}$. Thus, by inserting Eq. \eqref{nonresonance} into Eq.
\eqref{argument}, one has

\begin{equation}
\begin{array}{ll}
\omega_{i,\pm}(r)  &= \underbrace{\frac{1}{2 \pi i} \oint \limits_{\displaystyle \mathscr{C}} dz
\  z \ \partial_z  \ln \left[ z^2 -  \omega_{A,B}^2 \right]}_{\displaystyle \omega_{A,B}}  +  \\
& \hspace{4cm}  \frac{1}{2 \pi i} \oint \limits_{\displaystyle
 \mathscr{C}} dz
\  z \ \partial_z
\ln \left[ 1  \mp  \frac{ \zeta_A\zeta_B(\chi'_{ii}(r,z))^2}{ (z^2 -  \omega_{A,B}^2) \left(\omega_A^2 - \omega_B^2 \right) }
\right]. \\
\end{array}
\end{equation}

The second integral is computed by integration by parts. After Taylor expansion of the logarithm, one gets

\begin{equation}
\omega_{i,\pm}(r) \simeq  \omega_{A,B}  \pm  \frac{1}{2 \pi i} \oint \limits_{\displaystyle \mathscr{C}} dz
 \frac{ \zeta_A \zeta_B (\chi'_{ii}(r,z))^2}{ (z^2 -  \omega_{A,B}^2) \left(\omega_A^2 - \omega_B^2 \right)  }, 
\end{equation}

at lowest order. On the other hand, since the function
$\zeta_A\zeta_B(\chi'_{ii}(r,z))^2/ (z +  \omega_{A,B}) \left(\omega_A^2 - \omega_B^2 \right) $ is analytic everywhere
inside $\mathscr{C}$, we can make use of Cauchy's integral formula to find:

\begin{equation}\label{nonresonantshift}
\omega_{i,\pm}(r) \simeq  \omega_{A,B}  \pm
\underbrace{\frac{ \zeta_A \zeta_B(\chi'_{ii}(r,\omega_{A,B})))^2}{ 2 \omega_{A,B} \left(\omega_A^2 - \omega_B^2 \right)}}_{\displaystyle
 \Delta\omega_{A,B, i}(r)}.
\end{equation}

The normal frequencies thus found are equal to the frequencies of the dipole $\omega_{A,B}$
plus a shift due to the interaction. Note that $\omega_{A,B}$ corresponds
to $\omega_{i,\pm}(r)$ when $\boldsymbol{\mu}_A$ and  $\boldsymbol{\mu}_B$ are not interacting ($r\rightarrow \infty$).
Let us also note that if the contour $\mathscr{C}$ had been chosen on the left part of the complex plane (complex numbers
with negative real parts) when applying the argument principle, we would find the additive inverse of $\omega_{i,\pm}$ of Eq. \eqref{nonresonantshift}. Finally,
the dipole moments of the molecules, \textit{e.g.}, molecule $A$, can be given as:

$$
\mu_{A,i}(t) = \sum \limits_i \mu_{A,i,+}^{(1)}e^{i\omega_{i,+} t} + \mu_{A,i,+}^{(2)}e^{-i\omega_{i,+} t} +
\mu_{A,i,-}^{(1)}e^{i\omega_{i,-} t} + \mu_{A,i,-}^{(2)} e^{-i\omega_{i,-} t},
$$

\textit{i.e.}, a sum of six uncoupled oscillators with frequencies $\omega_{i,\pm}$, $i=1,2,3$ and mean energies $\omega_{i,\pm} J_{i,\pm}$
where $J_{i,\pm}$ are action constants depending on initial conditions. Therefore, the total average energy of the system is

\begin{equation}\label{nonresonant energy}
 \begin{array}{ll}
  E_{tot} &= \sum \limits_{i} E_{i,+} + E_{i,-} = \sum \limits_{i} \omega_{i,+} J_{i,+} + \omega_{i,-} J_{i,-} \vspace{0.3cm} \\
	  &=  \underbrace{\sum \limits_{i}\omega_{A} J_{i,+} + \omega_{B} J_{i,-}}_{ \substack{ \vspace{0.1cm} \\ \text{  \small Energy of the} \vspace{0.1cm} \\ \text{\small uncoupled system}  }}
+ \ \ \underbrace{\sum \limits_{i}\Delta\omega_{A, i}(r) J_{i,+} - \Delta\omega_{B, i}(r) J_{i,-}}_{ \substack{ \vspace{0.1cm} \\ \text{\small Interaction Energy \textit{U(r)}}  }}
 \end{array}
\end{equation}

\medskip

where the second sum is  the interaction energy $U(r)$ of the system which, according to Eq. \eqref{nonresonantshift}, scales linearly with
$\left(\chi_{ii}'(r,\omega_{A,B})\right)^2$ in the off-resonance case. Thus, as a consequence of  Eqs. \eqref{electricsus}, one gets

\begin{equation}
U(r)\propto \pm \left[\chi_{ii}'(r,\omega_{A,B})\right]^2 \sim\ \pm\frac{1}{r^6},
\end{equation}

in the limit $r \ll c/\omega_{A,B}$, \textit{i.e.}, the interaction potential $U(r)$ becomes short-range in the near zone limit. At very large intermolecular 
distance, ($r\gg c/\omega_{A,B}$, \textit{i.e.}, far zone limits) it oscillates  with a $1/r^2$ envelope.
Let us remark that $U(r)\propto \frac{1}{r^6}$ is a Van der Waals-like potential but not a true Van der Waals potential which stems from virtual photon
exchange, whereas our computation corresponds to a real exchange of electromagnetic energy.

\subsection{Resonant Case}

When $\omega_A \simeq \omega_B = \omega_0$, Eq. \eqref{secdegre} is simply reduced to

\begin{equation}\label{resonance0}
\underbrace{\omega_{i,\pm}^2 - \omega_0^2  \mp  \sqrt{\zeta_A\zeta_B}\chi'_{ii}(r,\omega_{i,\pm})}
_{\displaystyle \Theta_{i,\pm}(r,\omega_{i,\pm})} = 0.
\end{equation}

Similarly to the non-resonant case, we find that if $\Theta_{i,\pm}(r,z)$ is analytical
inside and on a contour $\mathscr{C}$ around $\omega_{i,\pm}$, and for which the inequality

\begin{equation}
\sqrt{\zeta_A\zeta_B}\chi'_{ii}(r,z)  <
\left| z^2 -  \omega_0^2 \right|,
\end{equation}

is valid for all $z$ on $\mathscr{C}$, then $\omega_0$ is also located inside the contour and the solution
$\omega_{i,\pm}(r)$ may be given by

\begin{equation}
\omega_{i,\pm}(r)  = \underbrace{\frac{1}{2 \pi i} \oint \limits_{\displaystyle \mathscr{C}} dz
\  z \ \partial_z  \ln \left[  z^2 -  \omega_0^2 \right]}_{\displaystyle \omega_0}  -   \frac{1}{2 \pi i} \oint \limits_{\displaystyle
 \mathscr{C}} dz  \ln \left[ 1  \mp  \frac{\sqrt{\zeta_A\zeta_B}\chi'_{ii}(r,z)}{ z^2 - \omega_0^2}.
\right].
\end{equation}

In comparison with Eq. \eqref{nonresonance}, Eq. \eqref{resonance0} is exact, so it makes sense
to write the complete series expansion of the logarithm :

\begin{equation}
\begin{array}{ll}
\omega_{i,\pm}(r) &=    \omega_0  +   \sum \limits_{n=1}^{\infty} \frac{(\pm 1)^n}{2 \pi in} \oint  \limits_{\displaystyle \mathscr{C}}  dz
  \left\{\frac{  \sqrt{\zeta_A\zeta_B}\chi'_{ii}(r,z)}{ z^2 -  \omega_0^2  } \right\}^n \\
&=    \omega_0  +   \sum \limits_{n=1}^{\infty} \frac{(\pm 1)^n}{2 \pi in} \oint  \limits_{\displaystyle \mathscr{C}}  dz
 \frac{1}{(z -  \omega_0)^n} \left\{\frac{  \sqrt{\zeta_A\zeta_B}\chi'_{ii}(r,z)}{ z +  \omega_0  } \right\}^n.
\end{array}
\end{equation}

Hence, Cauchy's integral formula may be applied at all orders to the function in brackets since
it is analytic inside and on $\mathscr{C}$ :

\begin{equation}\label{resonantshift}
\omega_{i,\pm}(r) =   \omega_0  + \sum \limits_{n=1}^{\infty} \frac{(\pm 1)^n}{n!}
\frac{d^{n-1}}{d\omega^{n-1}} \left[\left\{
\frac{\sqrt{\zeta_A\zeta_B}\chi'_{ii}(r,\omega)}{\omega + \omega_0} \right\}^n \right]_{\omega=\omega_0}.
\end{equation}

Using similar arguments, it may
be shown that any function $f$  analytic on and inside $\mathscr{C}$
may be expanded as a power series in $\omega_{i,\pm}(r)$ through the formula :

\begin{equation}\label{resonantfunc}
g(\omega_{i,\pm}(r)) =   g(\omega_0)  + \sum \limits_{n=1}^{\infty} \frac{(\pm 1)^n}{n!}
\frac{d^{n-1}}{d\omega^{n-1}}
\left[g'(\omega)
\left\{\frac{\sqrt{\zeta_A\zeta_B}\chi'_{ii}(r,\omega)}{ \omega + \omega_0  } \right\}^n \right]_{\omega=\omega_0} .
\end{equation}

Now, if we consider Eq. \eqref{resonantshift} at lowest order, one has:

\begin{equation}\label{resonantshiftbis}
\omega_{i,\pm}(r) \simeq \omega_0  \pm  \underbrace{\sqrt{\zeta_A\zeta_B} \frac{\chi'_{ii}(r,\omega_0)}{2 \omega_0}}_{\displaystyle
 \Delta\omega_{0,i}(r)}.
\end{equation}

The first contribution to the frequency shift is now proportional to $\chi'_{ii}$. In this case, the
total energy is:

\begin{equation}\label{resonant energy}
  E_{tot} =  \underbrace{\sum \limits_{i} \omega_0 J_{i,+} + \omega_0 J_{i,-}}_{ \substack{ \vspace{0.1cm} \\ \text{  \small Energy of the} \vspace{0.1cm}
\\ \text{\small uncoupled system}  }}
+ \ \ \underbrace{\sum \limits_{i}\Delta\omega_{0,i}(r)\left( J_{i,+} -  J_{i,-} \right)}_{ \substack{ \vspace{0.1cm} \\ \text{\small Interaction Energy \textit{U(r)}}  }}
\end{equation}

\bigskip

Thus, according to Eqs. \eqref{resonantshiftbis} and \eqref{electricsus}, $U(r)$ will be a polynomial in $1/r^\alpha$ with $\alpha \le 3$ (the dimension of physical space)
making the potential long-range at all distance. Hence, intermolecular electrodynamic interactions between dipoles
oscillating at similar frequencies are expected to have a much longer range of action with respect to off-resonance interactions.
More specifically, from Eqs. \eqref{electricsus} in the limit $r \ll c/\omega_{0}$ (near zone limit), and using Eqs. \eqref{resonant energy} and  \eqref{resonantshiftbis},
the interaction energy as a function of the intermolecular distance is

\begin{equation}
U(r)\propto \pm \chi_{ii}'(r,\omega_{A,B}) \sim\ \pm\frac{1}{r^3}\ 
\end{equation}

because the term  $1/r^3$ is dominant in the expression of  $\chi_{ii}'(r,\omega)$ while in the intermediate and far zone limits the dominant contribution is
a spatially oscillating one with a $1/r$ envelope (see the end of the Appendix for the functional form  of $\chi_{ii}'(r,\omega)$).

\newpage

\section{Numerical estimates of resonant/non-resonant interactions between biomolecules}\label{numerical estimates}

\subsection{Susceptibility and frequency shifts}

\begin{figure}
\begin{center}
    \includegraphics[width=4.0in]{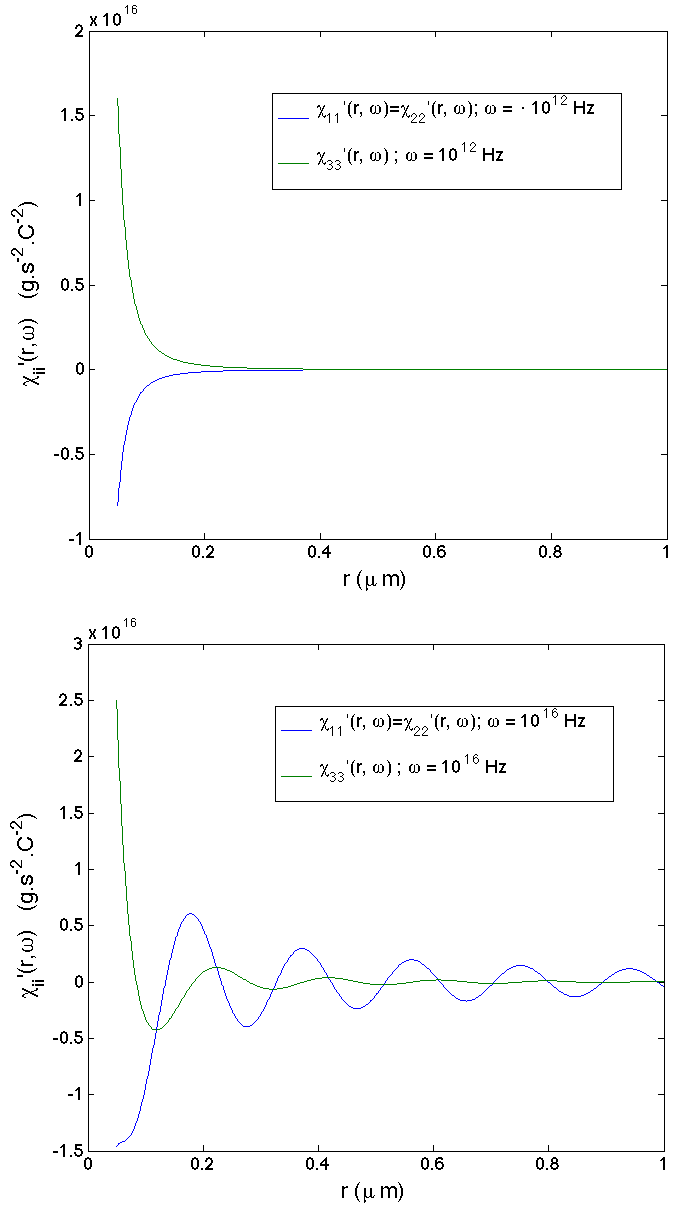}
\caption{(Color online) (a) Real part of the electric susceptibility matrix elements $\chi_{ii}'(r,\omega)$ computed from Eqs. \eqref{electricsus}
as a function of the distance $r$ (the index $E$ on $\chi_{ii}$ has been omitted to reduce the amount of notation).
The parameter values are $\varepsilon(\omega)=1$ and $\omega=10^{12} \ \text{Hz}$. (b) Same
as Figure (a) except that $\omega = 7.24 \cdot 10^{15} \ \text{Hz}$.}\label{susceptibilityplot}
\end{center}
 \end{figure}

To link up the above analytical results with the question of long-range molecular recognition in living matter, numerical estimation of resonance and off-resonance
interactions is carried out using parameter values related to standard biomolecules. In the introduction, we reported the existence of experimental
observations of low-frequency conformational oscillation modes in the Raman and far-infrared spectra of polar proteins \cite{proteins}. This is in line with our classical
description since quantum effects become relevant when typical frequencies exceed $\omega\sim k_BT/ \hbar=3.92.10^{13} ~ \text{Hz}$, at $T=300 ~ \text{K}$. For instance,
one can set the resonance frequency $\omega_0$ around $10^{11} -10^{12} ~ \text{Hz}$, as also suggested by Fr\"ohlich \cite{frohlich78,frohlich80}. To compute the coupling constants $\zeta = Q^2/m$, $10$ elementary charges and a mass of $m = 20$ kDa are taken as typical values from dipole moment and mass of small proteins (the same values are used for both molecules so that $\zeta_A=\zeta_B=\zeta$). As a first step in the estimation of the interaction potential $U(r)$, normal frequencies shifts are computed in both resonant and non-resonant cases. From equations  \eqref{nonresonantshift} and \eqref{resonantshiftbis}, one has:

\begin{equation}\label{shifts}
\left \{
\begin{array}{l}
 \Delta\omega_{A,B, i}(r)= \frac{ \zeta^2 (\chi'_{ii}(r,\omega_{A,B})))^2}{ 2 \omega_{A,B} \left(\omega_A^2 - \omega_B^2 \right)}, \  \  \ \text{when} \ \omega_A \gg \omega_B  \vspace{0.3cm} \\
\Delta\omega_{0,i}(r)= \zeta \frac{\chi'_{ii}(r,\omega_0)}{2 \omega_0} \  \  \ \text{when} \ \omega_A \simeq \omega_B = \omega_0, 
\end{array}
\right.
\end{equation}

where the diagonal elements of the susceptibility matrix $\chi_{ii}(r,\omega)$  are given by Eq. \eqref{electricsus} (the prime symbol stands for the real part). In Figure \ref{susceptibilityplot}, we have plotted the real parts of $\chi_{ii}(r,\omega)$ as a function of $r$ for two values of $\omega$, namely,  $\omega=10^{12} \ \text{Hz}$ and $\omega=10^{16} \ \text{Hz}$, below and above the classical/quantum limit, respectively. We are especially interested in $\chi_{ii}(r,\omega)$ when the intermolecular distance  $r$ is much larger than the dimensions of typical macromolecules, estimated around $5 \ \text{nm}$, but less than cellular dimensions $\sim 1 \ \mu\text{m}$. The dielectric constant $\varepsilon$ also appears in the expression of the susceptibility matrix as a function of the frequency. The dielectric constant of water is $80$ in the electrostatic limit, \textit{i.e.}, $\omega \rightarrow 0$ but for large enough frequencies, typically larger than a dozen of GHz \cite{hasted}, $\varepsilon$ drops to a few units, more exactly, $2\le  \varepsilon \le 10$. As a compromise, we will assume $\varepsilon(\omega) =4$ for all $\omega$ we will consider in the following. 
As mentioned above, the elements of the susceptibility matrix were also plotted at very large frequency $\omega=10^{16} \ \text{Hz}$ [Figure \ref{susceptibilityplot} (b)]. Even though quantum effects are important at such a frequency, quantum computations reveal that the "quantum" susceptibility matrix is the same as the classical one as well as its mathematical contribution to the interaction potential \cite{jthesis}. Very high frequencies might be relevant to account  for certain dynamical properties of DNA molecules since a marked peak is observed in their polar spectra at wavelengths of $2600 \ \mathring{\text{A}}$, which is equivalent to a frequency $\omega\sim7.10^{15} \ \text{Hz}$.  When $\omega=10^{12} \ \text{Hz}$, it is seen from Figure \ref{susceptibilityplot} (a) that the $\chi_{ii}'(r,\omega)$'s are essentially monotonic in $r$. The reason is that the intermolecular distance is much smaller than the characteristic wavelength $2\pi c/\omega$ of the electric field. From Eqs. \eqref{electricsus}, the susceptibility matrix elements are proportional to $1/r^3$ when $r\ll c/\omega$. In Figure \ref{susceptibilityplot} (b), the wavelength is small enough to observe oscillations at relatively low distances (less than $1 \ \mu \text{m}$). At $r=100 \ \text{nm}$, a frequency of at least $\omega_0=c/r=3.10^{13} \ \text{Hz}$ is needed to observe field retardation effects.

The frequency shifts are estimated from Eq. \eqref{shifts} when the intermolecular distance is set to $r=10\ \text{nm}$ and $r=100 \ \text{nm}$. With the parameter values specified above, the frequency shifts of the transverse modes ($i=1,2$) at resonance (with $\omega_0=10^{11} \ \text{Hz}$) are found equal to $8.63. 10^{8}\ \text{Hz}$ and $8.63. 10^{5} \ \text{Hz}$  at $r=10\ \text{nm}$ and $r=100 \ \text{nm}$, respectively. To estimate off-resonance frequency shifts, a detuning $(\omega_A-\omega_B)/\omega_B = 10\%$, with $\omega_B = 10^{11} \ \text{Hz}$, is used. Keeping the same values of the parameters,
the frequency shifts of the transverse modes ($i=1,2$) are found equal to $6.453. 10^{7} \ \text{Hz}$ and  $64.53 \ \text{Hz}$ at $r=10\ \text{nm}$ and $r=100 \ \text{nm}$, respectively.
Since the interaction potential is directly proportional to the frequency shifts in both resonant and non-resonant cases, it is found that  dipolar interactions at resonance may exceed by several orders of magnitude off-resonance ones, typically when the distances get large with respect to biomolecular scale length (around $10 - 50 \ \mathring{\text{A}}$).
This gap even widens by increasing the frequency of both dipoles.

\subsection{Interaction energy and equivalent dipole overcoming thermal noise}\label{numerical estimates interaction energy}

Let us now give some estimates related to the interaction energy. For the sake of clarity, we will restrict ourselves to the resonant case. As shown in section \ref{classical interactions}, the interaction potential $U(r)$ requires the value of the actions $J_{i,\pm}$ of each normal mode (see Eq. \eqref{resonant energy}). For a conservative system, actions are constant quantities which depend on initial conditions, \textit{i.e.}, on initial dipole moments and velocities of each molecule. However, bio-macromolecules are nonlinear systems that may strongly interact with their environment resulting in long-lived non-equilibrium stationary states. Energy supply could also be well provided by cellular machinery \cite{michal} (e.g., the energy released by ATP or GTP hydrolysis or the wasted energy released from mitochondria). Such conditions may lead to metastable states characterized by $J_{i,+}$ very different from $J_{i,-}$ whose values are not directly predictable from other physical quantities measured in equilibrium conditions (e.g., dipole moments of the molecules
as they are estimated \textit{in vitro} experiments). Hence, estimating the exact values of $J_{i,+}$ and $J_{i,-}$ turns out to be a non-trivial task as it requires further investigation on how  the molecules interact with their surrounding medium as well as a detailed description of the internal dynamics of the molecules, which is beyond the scope of this paper. 

Even though the interaction energy can not be estimated in an exact way, we can still compute it when the difference in the actions $\Delta J=J_{i,+}-J_{i,-}$ is of the order of 
thermal fluctuations, \textit{i.e.},  $\Delta J=k_BT/\omega_0$. This will provide a lower bound for the interaction potential. From Eq. \eqref{resonant energy}, the interaction energy becomes
$U(r)\sim \Delta\omega_{0,i}(r) k_B T/\omega_0$. Using the estimates of the frequency shifts given above, we find that $U(r)$ is equal to $3.57.10^{-16} \ \text{erg}$  and 
$3.57.10^{-19}\ \text{erg}$, at $r=10\ \text{nm}$ and $r=100 \ \text{nm}$, respectively. These values should be compared with the value of $kT=4.14.10^{-14} \ \text{erg}$ at $300 \ \text{K}$, which reveals
that $\Delta J$ should be at least 2 orders of magnitude larger than Boltzmann action so that dipole interactions balance exactly thermal energy at $r=10 \ \text{nm}$. To give an idea of what 2 orders
of magnitude mean in terms actions variables, we can compute the dipole moment equivalent to a given value of $\Delta J$. Writing 

$$
\omega_0 \Delta J = \frac{\left(\Delta p \right)^2}{2m} +\frac{1}{2} m\omega_0^2 {\left(\Delta x\right)}^2,
$$

where we assumed $m_A=m_B=m$ for simplicity, the dipole displacement $\Delta x$ is maximum when $\Delta p=0$, so that
\begin{equation} 
\Delta x_{max}= \sqrt{\frac{2\Delta J}{m\omega_0}}.
\end{equation}

Again, dipole interactions overcome thermal noise when $U(r)\sim k_BT$, that is, $\Delta J=k_BT/\Delta \omega_0(r)$ where $\Delta \omega_0(r)$ is given by Eq. \eqref{shifts} (the subscripts $i$ are omitted as, roughly speaking, each component has the same order of magnitude).  Using Eq. \eqref{shifts}, the maximum amplitude of the equivalent dipole is given by

\begin{equation} \label{dipole amplitude}
\Delta x_{max}= 2\sqrt{\frac{k_BT }{ Q^2 \chi'(r,\omega_0)}} \ \xrightarrow[r\ll c /\omega_0]{}  \ 2\sqrt{\frac{k_BT r^3  \varepsilon(\omega_0) }{Q^2}},
\end{equation}

\setlength{\tabcolsep}{10pt}
\setlength{\extrarowheight}{1pt}
\begin{table}
\footnotesize
   \centering
\begin{tabular}{g g g g c c c c} \toprule 
    $Z$\footnote{Z: number of charges with $Q=Ze$.} &  $m \ \text{(kDa)}$ & $\omega_0 \ (\text{THz})$\footnote{$\omega_0$: resonance frequency.} & $r \ (\text{nm}) \footnote{r: intermolecular distance.}$ &  $\Delta \omega_0(r) \ (\text{Hz})$\footnote{$\Delta \omega_0(r)$: frequency shift due to dipolar interactions. Computed from Eq. \eqref{shifts}.} & $\Delta x_{max} \ (\text{\AA})$\footnote{$\Delta x_{max}$: maximum amplitude that should carry any of the dipoles so that dipole interactions overcome thermal noise at separation $r$. Computed from Eq. \eqref{dipole amplitude}.} & $\mu \ (\text{D})$\footnote{$\mu=Q\Delta x_{max}$:  maximum dipole moment required to overcome thermal noise at $r$.} & $t_{1/2} \ (\text{s})$ \footnote{$t_{1/2}$: half-life associated with radiation losses. Computed from Eq. \eqref{halflife}.}  \\[0.3cm]  \colrule   
    $10$ & $10$ & $0.1$ & $10$ & $1.72 .10^9$ & 17.0 & $813.88$ & $0.733$\\ 
    $50$ & $10$ & $0.1$ & $10$ & $4.31.10^{10}$ & 3.39 & $813.88$ & $2.931 .10^{-2}$\\ 
    $10$ & $20$ & $0.1$ & $10$ & $8.63.10^8$ & 17.0 & $813.88$  & 1.466 \\
    $50$ & $100$ & $0.1$ & $10$ & $4.31.10^9$ & 3.39 & $813.88$ & $0.293$ \\ 
    $10$ & $10$ & $1$ & $100$ & $1.726.10^5$ & 535.8 & $25737.1$ & $7.329 .10^{-3}$ \\ 
    $50$ & $100$ & $1$ & $100$ & $4.31.10^5$ & 107.1 & $25737.1$  & $2.931 .10^{-3}$ \\[0.3cm]
     \botrule
\end{tabular}
\caption{\small Numerical estimates of physical quantities connected with the interaction energy of two oscillating dipoles at resonance ($\omega_A\simeq\omega_B=\omega_0)$. Grey columns correspond to
parameters that were initially fixed.  The temperature $T$ and the dielectric constant $\varepsilon$ were set to $300$ K and $4$, respectively, for each estimate. }
\label{estimates}
\end{table}

where we have used $\zeta=Q^2/m$ and, using Eq. \eqref{electricsus}, replace $\chi'(r,\omega_0)$ by its limit when $r\ll c /\omega_0$.  $\Delta x_{max}$ can be interpreted as the maximum amplitude that should exhibit any of the dipoles so that resonant interactions balance exactly thermal fluctuations at a given $r$. Using parameter values introduced above ($Q$ given by $10$ elementary charges, $\varepsilon(\omega_0)=4$), we find that $\Delta x_{max}= 1.694.10^{-7} \ \text{cm}  = 16.9 \ \text{\AA}$  at $r=10 \ \text{nm}$. The corresponding dipole moment is $\mu=Q\Delta x_{max}= 8.1388.10^{-16} \ \text{statC.cm}=813.88 \ \text{D}.$ The larger the value of $Q$, the smaller the value of the polar amplitude $\Delta x_{max}$. For instance, when $Q$ is given by $50$ elementary charges,  $\Delta x_{max}= 3.38905.10^{-8} \ \text{cm}$ $ = 3.38905 \ \text{\AA}$ at $r=10 \ \text{nm}$ while the equivalent dipole moment $\mu$ remains the same. Table \ref{estimates} gives a summary of the numerical estimates discussed so far plus other estimates obtained by varying parameters such as the intermolecular distance or the resonance frequency. In a recent paper \cite{nardecchia}, molecular dynamics simulations performed for an ensemble of interacting molecules reveal that $1/r^3$ resonant interactions might already have a profound influence on the diffusion dynamics when the interacting energy is equal to $kT/10$ at $r=100 \ \text{nm}$. Substituting $kT$ with $kT/10$ in Eq. \eqref{dipole amplitude}, we find that the equivalent dipole moment $\mu=Q\Delta x_{max}$ is equal  to $8138.8 \ \text{D}$ when $r=100 \ \text{nm}$. 
This value together with the values of the dipoles moments given in the table can appear to be very large in comparison with the dipole moments of a wide class of proteins given around a few hundred debyes \cite{takashima96, takashima02}. However, a mathematical approach similar to the one of section \ref{classical interactions} might be applied to estimate the interaction energy between two sets of oscillating dipoles instead of two single ones. The dynamical variables will be no longer the dipole moments of the molecules as we should take care of the local density of dipoles but the polarization fields, \textit{e.g.}, $P_A= n_A \langle \mu_A \rangle$, with $n_A$ the number of dipoles making up the "molecule" A  with an average dipole moment $\langle \mu_A \rangle$ (and similarly for the "molecule" B). Assuming that the dipoles of each "molecule" oscillate in phase, one can estimate the maximum amplitude that should carry any of the dipoles so that resonant interactions balance thermal noise at distance $r$. Similarly to the above computations, we can show that  $\Delta x_{max}= 2\sqrt{k_BT r^3  \varepsilon(\omega_0)/n^2 Q^2}$, where we have supposed that the number of dipoles composing each molecule was approximatively the same $(n_A\simeq  n_B=n)$. In this case, the average dipole amplitude is $n$ times less than the amplitude required by single dipoles, which might lead to more realistic values of the dipole moments when $n$ is large.  At the same time, the frequency shifts in the case of two interacting sets of dipoles should be computed according to Eq. \eqref{shifts} with coupling constants given by $\zeta=nQ^2/m$ instead of $Q^2/m$. A possible value for $n$ could be well given by the number of water molecules surrounding protein structures \cite{kabir}. It has been recently demonstrated by the work of G.Pollack \cite{pollack} that hydrophilic surfaces have long-range effects  on aqueous solutions containing solute molecules. The so-called "exclusion zone" (EZ) is created by repulsive forces extending over distances of many Debye lengths, possibly on the $\mu$m scale. Since proteins are typically structured with a hydrophobic core and a hydrophilic exterior surface, it is expected that similar EZ effects occur in the case of proteins and protein aggregations. One of the characteristics of the water molecules surrounding biomolecules such as proteins and DNA is the formation of several ordered layers of water surrounding a biomolecule \cite{giambasu}. These layers exhibit electrostatic ordering and also structural organization, possibly involving a hexatic lattice \cite{tuszynski04}. These phenomena may all be linked and can be interpreted in the spirit of the Mercedes-Benz model of water \cite{bennaim} whereby a charged (hydrophilic) surface such as that of a protein attracts the oppositely charged orbital of a water molecule leaving the remaining three orbitals free to explore rotational freedom in the plane approximately parallel to the charged surface of the protein. The so-formed hexatic arrangements of water orbitals resemble helicopter blades or the logo of Mercedes-Benz, hence the name of the model. This suggests a very interesting new phenomenon since this positionally localized water molecules possess a dipole moment that is free to rotate around the axis  roughly perpendicular to the protein surface. These rotating dipole moments should undergo precessional motion if there is a sufficiently strong electric field created, for example, by the charge distribution of the protein.
For instance, in the case of tubulin, its net dipole moment is of the order of $1000 - 5000$ D \cite{tuszynski05} depending on the tubulin isoform. This then leads to a dynamic picture of thousand of water molecules with their dipole moments precessing at a frequency in the $10^{11}$ Hz domain which is sensitive to the electric field generated by the biomolecular surface. The latter should also depend on the environmental conditions (pH, ion concentrations, salt concentration, and so on).
Moreover, conformational changes in the states of a protein still affect the magnitude and direction of the electric field that drives the dynamics of these precessing dipoles. Therefore, this could serve as a mechanism for selective generation of attractive or repulsive forces between interacting proteins (with their water of hydration). The electric field of the protein may additionally exhibit vibrational modes due to collective excitation modes, ${\bf E} = {\bf E}_0 + {\bf E}_1 e^{i\omega t}$. This, through coupling with dipole moments of the water molecules, will result in the slaving of water dynamics to the collective dynamics of the protein and generate coherent vibrational dynamics of dipole moments. This scenario offers a new and so-far unexplored possible role of water dynamics in living processes.

\smallskip

\subsection{Radiation losses}

The last column of table \ref{estimates} corresponds to the half-life $t_{1/2}$ of the typical dipoles that were considered so far, \textit{i.e.}, the time needed for radiation losses to halve their energy. To estimate $t_{1/2}$, the total instantaneous power $P$ radiated by one dipole is first computed using the Larmor formula \cite{jackson, panofsky}

\begin{equation}\label{larmor}
P = \frac{2}{3}  \frac{Q^2 |\ddot{\mathbf{x}}|^2}{c^3},
\end{equation}

where $Q= Ze$ and the acceleration is given by $\ddot{\mathbf{x}} = - \omega_0^2  \mathbf{x}$. The half-life of the dipole is

$$
t_{1/2} = \frac{E_0/2}{P},
$$

with $E_0$ the initial energy of the equivalent dipole  $\textstyle E_0 \simeq \frac{1}{2}m \omega_0^2 |\mathbf{x}|^2$. Hence,

\begin{equation}\label{halflife}
t_{1/2} = \frac{3}{8} \frac{m c^3}{ (Ze)^2  \omega_0^2}.
\end{equation}

In the range of frequencies $10^{11}$-$10^{12} \ \text{Hz}$, $t_{1/2}$ is estimated for a typical protein from several milliseconds to around one second (see table \ref{estimates}). These times should be compared to characteristic encounter times $\tau$ of two molecules $A$ and $B$ (ligand and receptor for example) in a solution of initial concentration $c$, where $c=1/r^3$ and $r$ is the range of intermolecular distances of interest. As discussed in the previous section, the range $r = 10$ - $100 \ \text{nm}$ is investigated which is equivalent to a range of concentration of $10^{-3}-10^{-6} \ \text{mol.L}^{-1}$, respectively. An upper bound for $\tau$ can be directly deduced from the association rate of brownian molecules $k_a$ in a solution which is given by the Smoluchovski formula \cite{firstpaper,chandrasekhar}
\begin{equation}\label{ksmolu}
k_a = 4 \pi R D,
\end{equation}

with $R=R_A+R_B$ the sum of the radii of molecules $A$ and $B$ that we suppose spherical and $D=D_A+D_B$ the sum of their diffusion coefficients; $D_{A,B} = kT/\gamma_{A,B}$. As usual, the friction coefficient $\gamma_{A,B}$ of each molecule can be approximated from Stokes law: $\gamma_{A,B} = 6\pi \eta R_{A,B}$ where $\eta\simeq 10^{-2} \ \text{g.cm}^{-1}\text{.s}^{-1}$ is the dynamic viscosity of water at $300$ K. We also set $R_A = R_B =1 \ \text{nm}$ as a typical radius for small proteins. The characteristic half-life of the A-B reaction in the case of Brownian molecules can be written as
$$
\tau = \frac{1}{k_a c}.
$$
Using Eq. \eqref{ksmolu} and the values of the parameters introduced above, we find that $\tau = 9.06 .10^{-8} \ \text{s}$ - $9.06.10^{-5} \ \text{s}$ when the concentration is equivalent to intermolecular distances of $r = 10$ - $100 \ \text{nm}$, respectively. Therefore, resonant dipolar interactions could influence the dynamics of biomolecular reactions as the characteristic decay times of the dipole moments ($t_{1/2}$ in table \ref{estimates}) are well slower than typical association times of biomolecular reactions.  The small values of the decay times are explained by the fact that the frequencies of the dipoles are small. When $\omega_0 =10^{15}$ Hz, $t_{1/2}$ is found of the order of $10^{-8}$-$10^{-9} \ \text{s}$ which is negligible with respect to the values of $\tau$ found above. As far as frequencies below the terahertz range are concerned, the decay times remain relatively slow. Hence, in case long-range dipolar interactions require energy supply to overcome thermal noise as discussed in the previous section, occasional energy kicks, rather than constant energy inputs, would be enough to maintain the large dipole moments needed. Among possible sources of energy, ultra-weak photons constitute ideal candidates.  At the cellular and sub-cellular level, living systems are known to emit ultra-weak endogeneous photons without the need for external excitation \cite{chang,cohen,devaraj,kobayashi1,kobayashi2,quickenden,takeda,vanwijk,yoon}. This is only dependent on the presence of metabolic activity. These electromagnetic excitations are produced via various biochemical reactions, but principally from bioluminescent radical recombination reactions involving the very numerous reactive oxygen and nitrogen species which results in a subsequent relaxation of excited states giving rise to photon emission. The oxidative phosphorylation metabolism taking place in the mitochondria of living cells and lipid peroxidation appear to be a primary source for this activity \cite{nakano,thar}. Oxidative phosphorylation is the most common form of energy production in dividing cells. Neurons also continuously produce photons during their ordinary metabolism \cite{isojima,kataoka}, and it has been shown in vivo that the intensity of photon emission from rat brain correlates well with cerebral energy metabolism, electrical activity, blood flow, and oxidative stress \cite{kobayashi2, kobayashi3}. Moreover, Sun et al. \cite{sun} demonstrated that ultraweak bioluminescent photons can propagate along neural fibers and can be considered a means of neural communication. Also of interest is the reported radical recombination within mitochondria can emit photons in the UV range required to excite the chromophoric network within microtubules \cite{voeikov}.

\newpage

\section{Comparison Between Classical and Quantum Electrodynamic Interactions}\label{quantum classical}

Now that the interaction energy of system \eqref{system} has been worked out, it is of interest to note that 
the interaction energies $U_{\pm}$ computed from Eq. (\ref{nonresonant energy}) or Eq. (\ref{resonant energy}) as

\begin{equation}\label{quantum shifts}
U_{\pm}(r) = \hbar \sum_{i} \left({\omega}_{i,\pm}(r) - \omega_{A,B}  \right),
\end{equation}

\textit{i.e.}, with the substitutions $J_{i,+}=\hbar$, $J_{i,-}=0$ or $J_{i,+}=0$, $J_{i,-}=\hbar$, respectively,  correspond exactly to the
interaction energy (due to \textit{real} photons) between two neutral atoms when one of them is in an excited state. 
This is fully consistent with the idea that atoms and the radiation field mediating the interaction is generally described as an
ensemble of coupled oscillators whereas normal modes and susceptibilities are the same for classical and quantum oscillators. However, despite this
remarkable analogy, of course, the origin of the interactions is totally different. Whereas atomic dipoles
in the QED framework are associated with electronic transitions, the classical computations mainly apply to dipole oscillations associated with conformational
vibrations of macromolecules. In what follows, the classical/quantum correspondence of electrodynamic interactions 
is given more explicitly in both non-resonant and resonant cases.

\subsection*{Off-resonance case}

Off-resonance, the normal frequencies are given by Eq. \eqref{nonresonantshift} so that $U_{\pm}$ reads as:

\begin{equation}\label{energy shift1}
U_+(r) =  \frac{\hbar}{2\omega_A}\frac{\zeta_{A}\zeta_{B}}{\omega_{A}^{\ 2} -\omega_B^2}\sum_i \left(\chi'_{ii}(r,\omega_A)\right)^2
\end{equation}

and

\begin{equation}\label{energy shift2}
U_-(r) =   \frac{\hbar}{2\omega_B}\frac{\zeta_{A}\zeta_{B}}{\omega_{B}^{\ 2} -\omega_A^2} \sum_i \left(\chi'_{ii}(r,\omega_B)\right)^2.
\end{equation}

Again, the components of the electric susceptibility $\chi_{ii}(r,\omega)$ are given by  Eq. \eqref{electricsus} but have not been made explicit here for the sake of clarity. Even if Eqs. \eqref{energy shift1} and \eqref{energy shift2} have been obtained classically, $U_\pm(r)$ is here equivalent to the energy shifts due to the interaction
between two atoms $A$ and $B$ with distinct transition frequencies $\omega_A \neq \omega_B$ when one of the atoms is in an excited state ($U_+(r)$ is the energy shift when the atom $A$ is excited
whereas $U_-(r)$ is when the atom $B$ is excited). The analogy is better seen writing $\alpha^{A,B}_{class.}(\omega)$ the classical polarizabilities of each dipole such that:

\begin{equation}\label{classical polarizability}
\alpha_{class.}^{A,B}(\omega) := \frac{\zeta_{A,B}}{\omega_{A,B}^{\ 2} -\omega^2}.
\end{equation}

In this case, $U_{\pm}(r)$ has exactly the same form as the quantum shifts computed, for example, by L. Gomberoff et al. (see Eq. (44) in Ref. \cite{mclone1}; see also \cite{mclone2}) in the context mentioned above.  $U_+$ is the energy shift associated with the (approximated) wave function $\vert\psi_+\rangle = |e_A, g_B \rangle$ whereas
$U_-$ is the energy shift associated with $\vert\psi_-\rangle = |g_A, e_B\rangle$ (here we have noted $e_A$ or $g_A$ when the atom $A$ is excited or is in
its ground state, respectively, the atoms $A$ and $B$ being uncoupled; and similarly for $B$). Finally, we should remark that the classical computations carried out 
above do not allow to reproduce the QED contribution in the energy shift due to virtual photons, \textit{i.e.}, which accounts for the interaction between the ground states of two atoms.
As shown explicitly in Ref. \cite{power},  such a contribution purely arises from vacuum fluctuations. A fully quantum description is then needed to derive the corresponding potential (see also Refs. \cite{casimir} for other derivations of van der Waals forces between two ground state atoms).




\subsection*{Resonant case}

At resonance, \textit{i.e.}, when $\omega_A \simeq \omega_B = \omega_0$, the normal frequencies are given by Eq. \eqref{resonantshiftbis} so that $U_{\pm}$ is simply:
\begin{equation}
U_\pm(r) =  \pm \frac{\hbar}{2\omega_0} \sqrt{\zeta_A \zeta_B} \sum_i \chi'_{ii}(r,\omega_0).
\end{equation}

Here, $U_\pm(r)$, with $\chi_{ii}(r, \omega)$ given from Eq. \eqref{electricsus}, is the classical equivalent of the interaction energy between two two-levels atoms in an excited state with a
common transition frequency $\omega_0$. The quantum result is usually given in terms of the off-diagonal elements of the dipole operator in the unperturbed state 
$\mu_{A}^{\ ge} = \langle g_A \vert \hat{\mu}_{A} \vert e_A \rangle$
and $\mu_{B}^{\ ge} = \langle g_B \vert \hat{\mu}_{B} \vert e_B \rangle$ (here we have supposed a condition of isotropy for the dipole moment of both atoms). In addition, let us remind that the \textit{quantum} polarizability $\alpha_{quant.}$ depends on these elements such that:

\begin{equation}\label{quantum polarizability}
\alpha^{A, B}_{quant.}(\omega) := \frac{2}{3\hbar} \frac{\omega_{A, B} \left(\mu_{A,B}^{\ ge}\right)^2 }{\omega_{A,B}^{\ 2} -\omega^2}.
\end{equation}

Thus comparing Eqs. \eqref{quantum polarizability} and Eqs. \eqref{classical polarizability} in the resonant case, one gets the following classical/quantum equivalence:

$$
\zeta_{A,B} \leadsto \frac{2}{3\hbar}\omega_0 (\mu_{A,B}^{\ ge})^2,
$$

with $\omega_A \simeq \omega_B = \omega_0$, so that the quantum version of $U_{\pm}$ can be easily derived:

$$
U_\pm(r) =  \underbrace{\pm \frac{\hbar}{2\omega_0} \sqrt{\zeta_A \zeta_B} \sum_i \chi'_{ii}(r,\omega_0)}_{\displaystyle \text{classical}}   \ \ \leadsto \ \ 
 \underbrace{\pm \frac{1}{3}\mu_{A}^{\ ge} \mu_{B}^{\ ge}\sum_i \chi'_{ii}(r,\omega_0)}_{\displaystyle \text{quantum}}.
$$

The last term corresponds exactly to the quantum energy shifts given for example by Eq. (5) in Ref. \cite{stephen} (see also Eq. (2.27) in Ref. \cite{mclachlan})
whose approximated eigenstates are

\begin{equation}\label{entangled states bis}
\begin{array}{c}
\vert\psi_+\rangle =\frac{1}{\sqrt{2}}\left[ \vert e_A,g_B\rangle - \vert g_A,e_B\rangle\right] \ \ \text{and} \vspace{0.3cm}\\
\vert\psi_-\rangle =\frac{1}{\sqrt{2}}\left[ \vert e_A,g_B\rangle + \vert g_A,e_B\rangle\right],
\end{array}
\end{equation}

for $U_+$ and $U-$, respectively. Again, let us emphasize that the interaction potential at resonance is of a much longer range than the off-resonance one. From
a biological point of view, such frequency-selective interactions could be of utmost importance during the approach of a molecule toward its
specific target as mentioned in the Introduction. On the other hand, quantum states given by Eqs. \eqref{entangled states bis},
as entangled (excitonic) states are fragile. In living matter, noisy cellular
environment could be sufficient to entail decoherence over long distances making long-range interactions between atoms not very probable in this case.
In this way, classical computations have the advantage of getting rid of the problem of quantum coherence, since, in this case,  dipole moments
of biomolecules are associated to classical conformational vibrations rather than electronic transitions. Again, this is in line with the 
experimental observations of low-frequency oscillations modes in the Raman and Far-Infrared spectra of polar proteins.
These spectral features are commonly  attributed to collective oscillation modes of the whole molecule (protein or DNA) or
of a substantial fraction of its atoms.

\section{Long-range electrodynamic interactions at thermal equilibrium}\label{thermo section}

Long-range interactions between two biological dipoles were first considered by Fr\"ohlich \cite{frohlich72,frohlich80}.
Fr\"ohlich emphasized, \textit{inter alia}, that long-range interactions may occur at resonance even though the system of dipoles is close to thermal
equilibrium. In a biological context this could be problematic as switching on and off long-range recruitment forces
seems more fit to explain activation and inhibition processes at work at the molecular level in living matter. 
To clarify this point, let us consider the dipoles $\boldsymbol{\mu}_A$ and $\boldsymbol{\mu}_B$ of system \eqref{system} in thermal
equilibrium and suppose that $\hbar \omega_{A,B}\ll k_B T$ so that classical effects are dominant (as discussed in section \ref{numerical estimates}, this
implies frequencies less than $k_BT/ \hbar=3.92.10^{13} ~ \text{Hz}$ at $T=300 ~ \text{K}$, which is in agreement with experimental evidences of marked peaks in the vibration spectra of many proteins \cite{proteins}). For a system interacting with a thermal bath, the interaction energy is best described by the difference of free energies between the interacting system and the non-interacting one:  $\textstyle{U(r) =  F(r) - F(\infty) = -k_B T \ln \left[Z(r)/Z(\infty) \right]}$,
where $Z(r)$ is the partition function of the system when the dipoles are separated by a distance $r$. In thermal equilibrium, the partition function is computed from Boltzmann weights. This can be easily calculated in the space of the normal modes as those are equivalent to uncoupled harmonic oscillators:

\begin{equation}
\begin{array}{ll}
Z(r) & = \  \prod \limits_{i=1}^{3} \idotsint d\pi_{i,+} d\pi_{i,-} d\mu_{i,+} d\mu_{i,-}  \\
& \hspace{2cm} \exp \left [ - \frac{\pi_{i,+}^2 +
{\omega}_{i,+}^{\ 2}(r)\mu_{i,+}^2}{ 2 k_B T} \right ] \exp \left [- \frac{\pi_{i,-}^2 +
{\omega}_{i,-}^{\ 2}(r)\mu_{i,-}^2}{2k_BT} \right ] \\
& =   \prod \limits_{i=1}^{3} \frac{(2 \pi k_B T)^2}{{\omega}_{i,+}(r){\omega}_{i,-}(r)},
\end{array}
\end{equation}

where $\pi_{i,\pm}$ and $\mu_{i,\pm}$ stand for the components of normal coordinates related to momenta and positions, respectively. Possible non-linear contributions of the potential due to dipole anharmonicities, as mentioned at the beginning of section \ref{classical interactions}, have been omitted as they are supposed to be negligible compared with the harmonic part of the potential.

Thus the free energy difference is given by (see also Ref. \cite{frohlich80}):

\begin{equation}\label{free_energy}
U(r) =  k_BT \sum \limits_i \ln \left[ \frac{{\omega}_{i,+}(r) {\omega}_{i,-}(r)}{\omega_A \omega_B} \right],
\end{equation}

with $\lim \limits_{r \rightarrow + \infty} \omega_{i,\pm}(r) = \omega_{A,B}$, $\forall i$. To derive Fr\"ohlich results, we consider the
resonant case $\omega_A \simeq  \omega_B = \omega_0$, and substitute ${\omega}_{i,\pm}(r)$ in Eq. \eqref{free_energy} with their \textit{implicit} form (Eq. \eqref{resonance0}). At long
distances, we can use Taylor expansion: $\ln (1+x) \sim x$. One gets the formula of the interaction energy given by Fr\"ohlich by making explicit
the susceptibility matrix elements $\chi_{ii}$ from Eq. \eqref{electricsus} when $r \ll  c / \omega_0$. One obtains

\begin{equation}\label{frohlich_energy}
U(r)  \simeq  \frac{k_BT \sqrt{\zeta_A\zeta_B}}{2\omega_0^2}   \frac{1}{ r^3} \sum \limits_i
\sigma_i \left\{ \frac{1}{\varepsilon(\omega_{i,+})} - \frac{1}{\varepsilon(\omega_{i,-})} \right\},
\end{equation}

with $\sigma_1, \sigma_2 = -1$, $\sigma_3 = 2$. Here the \textit{a priori} non-cancellation of the term in brackets was 
emphasized by Fr\"ohlich as involving long-range $1/r^3$ interactions between the dipoles even when the system is in thermal equilibrium. However, this form of $U(r)$
arises simply as the implicitness has not been solved yet. By using the \textit{Lagrange inversion theorem} (we substitute $g$
of Eq. \eqref{resonantfunc} with $1/\varepsilon$ and report the formula in Eq. \eqref{frohlich_energy}), one gets immediately that the term in brackets
in Eq. \eqref{frohlich_energy} vanishes at first order. Expansion of the Lagrange theorem to second order shows that this term goes as
$1/r^3$, making $U$ proportional to $1/r^6$, \textit{i.e.}, a short-range contribution.
Fr\"ohlich's statement on the existence of \textit{long-range} resonant interactions even at thermal equilibrium is thus
\textit{incorrect}. To compute the complete form of $U$ in thermal equilibrium, one can start from Eq. \eqref{free_energy} and use the \textit{explicit}
form of $\omega_{i,\pm}(r)$ derived above [Eq. \eqref{resonantshift}]:

$$
\begin{array}{ll}
U(r) & = k_BT \sum_i \ln \left[ \left( 1 + \frac{\sqrt{\zeta_A \zeta_B}}{2\omega_0^2} \chi_{ii}'(r,\omega_0) +
\frac{\zeta_A \zeta_B}{\omega_0} \frac{d}{d\omega} \left[ \left( \frac{\chi_{ii}'(r,\omega_0)}{\omega + \omega_0} \right)^2 \right]_{\omega_0} \right) \cdot \right.
\vspace{0.4cm} \\
& \hspace{4cm} \left. \left( 1 - \frac{\sqrt{\zeta_A \zeta_B}}{2\omega_0^2} \chi_{ii}'(r,\omega_0) +
\frac{\zeta_A \zeta_B}{\omega_0} \frac{\partial}{\partial\omega} \left[ \left( \frac{\chi_{ii}'(r,\omega_0)}{\omega + \omega_0} \right)^2 \right]_{\omega_0} \right)
\right] \vspace{0.4cm} \\
& = k_BT \sum_i \ln \left[  1 - \frac{\zeta_A \zeta_B}{4\omega_0^4} \left(\chi_{ii}'(r,\omega_0)\right)^2 + \frac{\zeta_A \zeta_B}{\omega_0}
\left( \frac{\partial\chi_{ii}'(r,\omega_0)}{\partial\omega}\frac{ \chi_{ii}'(r,\omega_0)}{2\omega_0^2} - \frac{\left(\chi_{ii}'(r,\omega_0)\right)^2}{4\omega_0^3}
\right)\right].
\end{array}
$$

Using the relation $\ln (1+x) \sim x$ valid at large $r$, the interaction potential becomes:

$$
U(r)  \simeq \frac{k_BT}{2\omega_0^4} \zeta_A \zeta_B \sum_i\left(\omega_0\frac{\partial\chi_{ii}'(r,\omega_0)}{\partial\omega} \chi_{ii}'(r,\omega_0)-
\left(\chi_{ii}'(r,\omega_0)\right)^2  \right).
$$

In particular, in the limit $r\ll c/\omega_0$, we have from Eqs. \eqref{electricsus}:

$$
\left\{
\begin{array}{l}
\chi_{ii}'(r,\omega_0) = \frac{\sigma_i}{\varepsilon(\omega_0)r^3} \vspace{0.3cm} \\
\frac{\partial\chi_{ii}'(r,\omega_0)}{\partial\omega} = \frac{\sigma_i}{r^3} \left( -\frac{1}{[\varepsilon(\omega_0)]^{\hspace{0.1mm} 2}}  \frac{d\varepsilon(\omega_0)}{d\omega_0} \right).
\end{array}
\right.
$$

with $\sigma_1, \sigma_2 = -1$, $\sigma_3 = 2$. The interaction potential $U(r)$ reads as:

\begin{equation}
U(r) =  - \frac{3 k_BT \zeta_A\zeta_B}{\omega_0^{\hspace{0.7mm}4}[\varepsilon(\omega_0)]^{\hspace{0.1mm} 2}}   \frac{1}{ r^6}
\left\{ 1 + \omega_0   \frac{d\ln[\varepsilon(\omega_0)]}{d\omega_0} \right\}.
\end{equation}

Here, the short-range nature of the potential is due to the use of Boltzmann distribution
which equally weights the normal modes energies. The cancellation of equal long-range contributions with opposite sign
[see Eq. \eqref{resonantshiftbis}] follows. Despite resonance, we conclude that electrodynamic interactions would have negligible effects with respect to Brownian motion, so that electrodynamic interactions
at thermal equilibrium are not expected to play a significant role in the dynamics of biomolecules.

\section{Concluding remarks}

In this paper, long-range electrodynamic interactions (EDI) between molecular systems have been investigated within a classical framework.  Our prime motivation has to do with biomolecular dynamics in a cellular environment.  Potential functions, \textit{i.e.}, force fields, used in standard molecular dynamics software packages \cite{gromacs, amber} usually involve short-distance two-body interaction potentials -- \textit{i.e.}, which have minimum influence beyond the Debye length -- including screened electrostatic Coulomb forces whose short-range nature is explained by the large amount of ionic entities located in the cell. However, Debye screening only applies for static charges as it becomes inefficient when electric fields with large enough frequency are involved, as was shown by Xammar Oro \cite{xammar}. Since proteins and  DNA/RNA molecules are characterized by high-frequency vibrational motions in the terahertz domain or above \cite{proteins,nucleotides}, it is worth investigating how forces of electrodynamic nature may influence the dynamics of biomolecules, especially over long distances. EDI are well known in QED whereas almost no litterature is available on classical interactions. In this paper, we reported that classical EDI show similar properties as quantum interactions in the dipole limit, \textit{i.e.}, at distances much larger than the dimensions of the molecules involved. Whenever the dipole moments of the molecules oscillate with the same frequency, long-range resonance interactions in $1/r^3$ are activated. Non-resonant conditions lead to short-range interactions in $1/r^6$. Numerical estimates regarding resonance and off-resonance EDI, \textit{e.g.}, the normal frequency shifts or the equivalent dipoles, were provided in section \ref{numerical estimates}. Even though the dipole moments needed to overcome thermal noise at large distance could appear large compared to dipole moments of small standard proteins, EDI can be well estimated for two collections of dipoles interacting upon each other, which lead to more realistic values of the dipoles moments involved. Water ordering around protein structures was suggested as a possible mechanism to enhance resonant EDI. In section \ref{quantum classical}, comparison between classical and quantum EDI was made. In section \ref{thermo section}, we emphasize why resonant interactions between biomolecules need nonequilibrium to be effective at a long distance, \textit{i.e.}, the excitation of one normal mode of the interacting system should be statistically ``favored'' (far beyond Boltzmann fluctuations) compared with other(s) modes. The same conclusion was reached by Tuszynski et al. \cite{tuszynski89} from numerical estimates. In section \ref{classical interactions}, normal modes have been computed from equations of motion \eqref{system0}  omitting anharmonic contributions, dissipative effects as well as possible external excitations of the oscillating dipoles. Dipole anharmonicities would give rise to nonlinear interactions in normal coordinates. In this case, one might expect long-lived non-equilibrium  states lacking energy equipartition among normal modes as, for example, in the case of nonlinearly coupled harmonic oscillators \cite{pettinibook}. If so, in a biological context, metabolic energy supply could be of utmost importance to maintain a high degree of excitation in a specific mode despite energy losses. This scenario was originally suggested by Fr\"ohlich \cite{frohlich68} who proposed a dynamical model to account for such non-thermal excitations in biological systems. In particular, he showed that a set of coupled normal modes can undergo a condensation phenomenon characterized by the emerging of the mode of \textit{lowest} frequency containing, in the average, nearly all the energy supply  \cite{frohlich68}. In the case of two interacting molecules, such a process could ensure the action constant of the mode of lowest frequency to be much greater than the action constant(s) related to other mode(s). Of course, this would result in an effective attractive potential whose amplitude is dependent on the ``stored'' energy. Finally, it is worth noting that retardations effects at large $r$ bring about interactions with a $1/r$ dependence [last terms in Eqs. \eqref{electricsus}],\textit{i.e.}, of much longer range with respect to the interactions proposed by Fr\"ohlich. This last result could be relevant for a deeper understanding of the highly organized molecular machinery in living matter, as emphasized in the Introduction.

\section{Appendix : Electromagnetic Field Generated by a Time-varying Source in a Medium} \label{EMF and source}

For the sake of clarity we outline below how the derivation proceeds of Eqs. \eqref{electricsus}. 
The starting point is the D'Alembert  wave equation  for the vector \text{potential $\boldsymbol{A}$} in Fourier space (in Lorenz gauge):

\begin{equation}\label{Fourier wave equation}
 \left[ k^2 - \frac{\omega^2}{c^2} \varepsilon(\omega) \right] \boldsymbol{A}(\boldsymbol{k}, \omega) = \frac{4\pi}{c}
\boldsymbol{J}(\boldsymbol{k}, \omega)
\end{equation}

Here, the dielectric constant $\varepsilon(\omega)$ is simply due to the
constitutive relation linking the displacement field $\boldsymbol{D}$ and the macroscopic electric field $\boldsymbol{E}$ in a homogeneous
isotropic dielectric medium: $\boldsymbol{D}(\boldsymbol{k},\omega) = \varepsilon(\omega) \boldsymbol{E}(\boldsymbol{k},\omega)$. 
From Eq. \eqref{Fourier wave equation}, $\boldsymbol{A}(\boldsymbol{r}, \omega)$ can be computed by inverse Fourier transform:

$$
\begin{array}{ll}
\boldsymbol{A}(\boldsymbol{r}, \omega) & = \frac{4\pi}{c}\left[ \frac{1}{(2 \pi)^3}  \int d^3 \boldsymbol{k}
\frac{\boldsymbol{J}(\boldsymbol{k}, \omega) }{k^2 - \omega^2 \varepsilon(\omega)/c^2 } e^{i\boldsymbol{k}\cdot \boldsymbol{r}} \right]\ ,
\end{array}
$$
and by convoluting with the inverse Fourier transform of $\boldsymbol{J} $ we get 

\begin{equation}\label{prod de conv}
\begin{array}{ll}
\boldsymbol{A}(\boldsymbol{r}, \omega) & = \frac{4\pi}{c} \int d^3 \boldsymbol{r'} \left[\frac{1}{(2 \pi)^3}\int d^3 \boldsymbol{k}
\frac{e^{i\boldsymbol{k}\cdot (\boldsymbol{r}-\boldsymbol{r'}) }}{k^2 - \omega^2 \varepsilon(\omega)/c^2 } \right]\boldsymbol{J}(\boldsymbol{r'}, \omega).
\end{array}
\end{equation}
The kernel is computed using spherical coordinates and integrating over the angles:

$$
 \begin{array}{ll}
 I(\boldsymbol{r},\omega) &\equiv  \frac{1}{(2 \pi)^3} \int d^3 \boldsymbol{k}
\frac{e^{i\boldsymbol{k}\cdot \boldsymbol{r} }}{k^2 - \omega^2 \varepsilon(\omega)/c^2 } \vspace{2mm} \\
&  =
\frac{1}{(2 \pi)^3} \int_0^{ \infty} k^2dk \int_ 0^\pi \sin(\theta) d\theta \int _0^{2\pi} d\varphi
\frac{e^{i k \cos(\theta) r }}{k^2 - \omega^2 \varepsilon(\omega)/c^2 },
\end{array}
$$
where, without loss of generality, the z-axis of the Cartesian coordinate system has been taken
along $\boldsymbol{r}$, so that $\boldsymbol{k}\cdot \boldsymbol{r} = k \cos(\theta) r$. We
show that $I(\boldsymbol{r},\omega)$ satisfy

$$
 I(\boldsymbol{r},\omega)  =
\frac{- i}{(2 \pi)^2 r} \int_0^{ \infty} dk \frac{ k }{k^2 - \omega^2 \varepsilon(\omega)/c^2 }
\left[e^{i k  r }  - e^{-i k  r }\right].
$$
In order to use the Residue Theorem, $k$ is substituted with $-k$ in the second term of the integrand to give 

$$
 I(\boldsymbol{r},\omega)  =
\frac{- i}{(2 \pi)^2 r} \int_{- \infty}^{ \infty} dk \frac{ k }{k^2 - \omega^2 \varepsilon(\omega)/c^2 } e^{i k  r }.
$$
Then the function

$$
f(z) = z e^{izr} \left(z^2 -\frac{\omega^2}{c^2} \varepsilon(\omega) \right)^{-1},
$$
is integrated on the semicircular contour of the upper complex plane which includes
the real axis. In this case,  $|z f(z)|$ approaches zero as $|z|$ approaches $+\infty$. 
Of the two simple poles of $f$ at $z = \pm z_0 = \pm \omega \sqrt{\varepsilon(\omega)}/c$ 
only one of them is located inside the contour.  When $\text{Im} (\omega \sqrt{\varepsilon(\omega)})>0$, the pole will
be $+z_0$, and vice-versa.  Hence, $I$ is simply found to be 

\begin{equation}
 I(\boldsymbol{r},\omega)  = \frac{1}{4\pi r} e^{\pm i \omega \sqrt{\varepsilon(\omega)} r /c},
\end{equation}

when $\text{Im} \left( \omega \sqrt{\varepsilon(\omega)} \right)$ is positive or negative, respectively, and Eq. \eqref{prod de conv} yields

\begin{equation}
\begin{array}{ll}
\boldsymbol{A}(\boldsymbol{r}, \omega) & = \frac{1}{c} \int d^3 \boldsymbol{r'}
\frac{e^{ \pm i \omega \sqrt{\varepsilon(\omega)} || \boldsymbol{r} - \boldsymbol{r'} || /c}}{|| \boldsymbol{r} - \boldsymbol{r'} ||} \boldsymbol{J}(\boldsymbol{r'}, \omega).
\end{array}
\end{equation}

Assuming the current $\boldsymbol{J}(\boldsymbol{r'}, \omega)$ is due to a molecule whose center of mass is
far from where the field is measured, one can take $r \gg r'$ and  $||\boldsymbol{r} - \boldsymbol{r}'||$ can be approximated as

$$
||\boldsymbol{r} - \boldsymbol{r}'|| =
r \sqrt{1 -\frac{2 \ \boldsymbol{r}'\cdot \boldsymbol{r}}{r^2} +
\left(\frac{r'}{r} \right)^2}  \simeq r
- \boldsymbol{r}'\cdot \boldsymbol{n} + \cdots
$$

where we have let $\boldsymbol{n}
\equiv \boldsymbol{r}/r$. Thus:

\begin{equation}\label{first app}
e^{ \pm i \omega \sqrt{\varepsilon(\omega)} || \boldsymbol{r} - \boldsymbol{r'} || /c}
\simeq e^{ \pm i \omega \sqrt{\varepsilon(\omega)}r /c} \left( 1 \mp \frac{ i\omega \sqrt{\varepsilon(\omega)}}{c} \ \boldsymbol{r}'\cdot \boldsymbol{n} + \cdots  \right).
\end{equation}
Likewise, one has

\begin{equation}\label{sec app}
\frac{1}{||\boldsymbol{r} - \boldsymbol{r}'||}=
\frac{1}{r} \left(1 -\frac{2 \ \boldsymbol{r}'\cdot \boldsymbol{r}}{r^2} +
\left(\frac{r'}{r} \right)^2 \right)^{-1/2}  \simeq  \frac{1}{r} \left( 1 + \boldsymbol{r}'\cdot \boldsymbol{n} + \cdots \right).
\end{equation}

\subsection*{Dipole approximation}

Considering only the first terms of the expansion in Eqs \eqref{first app} and \eqref{sec app},
as a first approximation, one gets:

$$
\begin{array}{ll}
\boldsymbol{A}(\boldsymbol{r},\omega) & \simeq \frac{e^{\pm i \omega \sqrt{\varepsilon(\omega)} r /c}}{cr}\int \limits
d^3 \boldsymbol{r}' \boldsymbol{J}(\boldsymbol{r}',\omega) \vspace{0.3cm} \\
&\simeq - \frac{e^{\pm i \omega \sqrt{\varepsilon(\omega)} r /c}}{cr}
\int  d^3 \boldsymbol{r}' \ \boldsymbol{r}' \left( \boldsymbol{\nabla} \cdot \boldsymbol{J} (\boldsymbol{r}',\omega) \right).
\end{array}
$$
Making use of the continuity equation, one finds

\begin{equation}
\boldsymbol{A}(\boldsymbol{r},\omega) = - \frac{ i\omega}{cr} e^{\pm i \omega \sqrt{\varepsilon(\omega)} r /c} \ \boldsymbol{\mu}(\omega),
\end{equation}
where $\boldsymbol{\mu}(\omega) = \int \limits d^3 \boldsymbol{r}'\boldsymbol{r}' \rho (\boldsymbol{r}',\omega)$ is the
(macroscopic) dipole moment associated with the distribution of charge $\rho$ of a molecule. Using (macroscopic) Maxwell equations
$\boldsymbol{B} = \boldsymbol{\nabla} \times \boldsymbol{A} \ \ \text{and} \ \
\boldsymbol{\nabla} \times \boldsymbol{B} = \boldsymbol{J} -
i \omega \varepsilon(\omega) \boldsymbol{E}/c$, we can easily find the magnetic and electric components of the radiation field:

\begin{equation}\label{magnetic}
\boldsymbol{B}(\boldsymbol{r},\omega)  = \frac{ i \omega \sqrt{\varepsilon(\omega)}}{c}
\cdot \frac{e^{\pm i \omega \sqrt{\varepsilon(\omega)} r /c}}{r^2}  (\boldsymbol{n} \times \boldsymbol{\mu} (\omega))  \left( 1 \mp \frac{ i\omega
\sqrt{\epsilon(\omega)} r}{c} \right)
\end{equation}

\begin{equation}\label{electric}
\begin{array}{ll}
\boldsymbol{E}(\boldsymbol{r},\omega) & =  - \frac{ ic }{\omega \varepsilon(\omega)}\boldsymbol{J}(\boldsymbol{r},\omega) -
\frac{e^{\pm i \omega \sqrt{\varepsilon(\omega)} r /c}}{\varepsilon(\omega) r^3}
\left\{ \left(\boldsymbol{\mu}(\omega) - 3 \boldsymbol{n} (\boldsymbol{n} \cdot \boldsymbol{\mu} (\omega)) \right) \left( 1 \mp \frac{ i\omega
\sqrt{\varepsilon(\omega)} r}{c} \right) - \right. \vspace{3mm} \\
& \left. \hspace{5cm} - \left(\boldsymbol{\mu}(\omega) -  \boldsymbol{n} (\boldsymbol{n} \cdot \boldsymbol{\mu} (\omega)) \right)  \frac{\omega^2
\varepsilon(\omega) r^2}{c^2}  \right\}.
\end{array}
\end{equation}

Again, let us stress that these equations are valid for $r$ large with respect
to the size of the molecule. Hence, we can assume $\boldsymbol{J}(\boldsymbol{r},\omega)=0$ in Eq. \eqref{electric} and write the electromagnetic field
in the compact form

\begin{equation}\label{electromag}
\begin{array}{ll}
\boldsymbol{B}(\boldsymbol{r},\omega) & = \boldsymbol{\chi}^B(r,\omega)\boldsymbol{\mu}(\omega) \vspace{0.3cm} \\
\boldsymbol{E}(\boldsymbol{r},\omega) & = \boldsymbol{\chi}^E(r,\omega)\boldsymbol{\mu}(\omega) ,
\end{array}
\end{equation}

where $\boldsymbol{\chi}^B(\boldsymbol{r},\omega)$ and $\boldsymbol{\chi}^E(\boldsymbol{r},\omega)$ are the susceptibility matrix of
the magnetic and electric fields. In particular, $\boldsymbol{\chi}^B(\boldsymbol{r},\omega)$ and $\boldsymbol{\chi}^E(\boldsymbol{r},\omega)$
 can be given in a diagonal form by taking the $z$ axis along $\boldsymbol{n}$. In this case:

\begin{equation}
 \begin{array}{rl}
\chi^B_{12}(\boldsymbol{r},\omega) = \chi^B_{21}(\boldsymbol{r},\omega) & =

\frac{i \omega \sqrt{\varepsilon(\omega)}}{c}
\cdot \frac{e^{\pm i \omega \sqrt{\varepsilon(\omega)} r /c}}{r^2} \left( 1 \mp \frac{ i\omega
\sqrt{\epsilon(\omega)} r}{c} \right)
,  \  \vspace{0.3cm} \\
\ \text{and}  \ \ \ \chi^B_{ij}(\boldsymbol{r},\omega) & =  0 \ \ \  \text{elsewhere},
\end{array}
\end{equation}

\bigskip

regarding the magnetic field, and :

\begin{equation}\label{electricsus0}
\begin{array}{rl}
\chi^E_{11}(\boldsymbol{r},\omega) = \chi^E_{22}(\boldsymbol{r},\omega) & = - \frac{e^{\pm i \omega \sqrt{\varepsilon(\omega)} r /c}}{\varepsilon(\omega) r^3}
\left(  1 \mp  \frac{ i\omega
\sqrt{\varepsilon(\omega)} r}{c}  -    \frac{\omega^2
\varepsilon(\omega) r^2}{c^2}  \right), \vspace{0.3cm} \\
\chi^E_{33}(\boldsymbol{r},\omega) & =  \frac{2e^{\pm i \omega \sqrt{\varepsilon(\omega)} r /c}}{\varepsilon(\omega) r^3}
\left(  1 \mp \frac{ i\omega
\sqrt{\varepsilon(\omega)} r}{c}   \right), \ \text{and} \  \vspace{0.3cm} \\
\chi^E_{ij}(\boldsymbol{r},\omega) & =  0 \ \ \  \text{for} \ \  i \neq j,
\end{array}
\end{equation}

\medskip

regarding the electric field.

\medskip

Equation \eqref{electricsus0} is the same of  Eq. \eqref{electricsus} of Section \ref{classical interactions}
where we considered two harmonic dipoles $A$ and $B$, so that  $\boldsymbol{\mu}$
in Eq. \eqref{electromag} should be replaced by $\boldsymbol{\mu}(\omega)= \boldsymbol{\mu}_{A,B} \delta(\omega - \omega_N)$.

\end{document}